\documentclass[english]{aa}
\usepackage{mathptmx}
\usepackage[T1]{fontenc}
\usepackage[latin9]{inputenc}
\usepackage{verbatim}
\usepackage{amsmath}
\usepackage{graphicx}

\makeatletter

\DeclareRobustCommand*{\lyxarrow}{%
\@ifstar
{\leavevmode\,$\triangleleft$\,\allowbreak}
{\leavevmode\,$\triangleright$\,\allowbreak}}

\makeatother

\usepackage{babel}

\begin{document}

\title{Black holes and galactic density cusps}

\subtitle{spherically symmetric anisotropic cusps}

\author{M. Le Delliou\inst{1}\and R.N. Henriksen\inst{2}\and J.D. MacMillan\inst{3}}

\institute{Instituto de Física Teórica UAM/CSIC, Facultad de Ciencias, C-XI,
Universidad Autónoma de Madrid\\
 Cantoblanco, 28049 Madrid SPAIN\\
\email{Morgan.LeDelliou@uam.es}\and Queen's University, Kingston,
Ontario, Canada \\
\email{henriksn@astro.queensu.ca}\and Faculty of Science, University
of Ontario Institute of Technology, Oshawa, Ontario, Canada L1H 7K4\\
\email{joseph.macmillan@gmail.com}}

\offprints{MLeD\hfill{}Preprint: IFT-UAM/CSIC-09-27}

\date{Submitted ...; Received ...; Accepted...}

\date{}

\abstract{
} {
In this paper we study density cusps that may contain central black
holes. The actual co-eval self-similar growth would not distinguish
between the central object and the surroundings. } {
To study the environment of a growing black hole we seek descriptions
of steady `cusps' that may contain a black hole and that retain at
least a memory of self-similarity. We refer to the environment in
brief as the `bulge' and on smaller scales, the `halo'.
 } {
We find simple descriptions of the simulations of 
 collisionless matter 
by comparing predicted densities, velocity dispersions and distribution
functions with the simulations. 
In some cases central point masses may be included by iteration. We
emphasize that the co-eval self-similar growth allows an explanation
of the black hole bulge mass correlation between approximately similar
collisionless systems.} {
We have derived our results from first principles assuming adiabatic
self-similarity and either self-similar virialisation or normal steady
virialisation. We conclude that distribution functions that retain
a memory of self-similar evolution provide an understanding of collisionless
systems. The implied energy relaxation of the collisionless matter
is due to the time dependence. Phase mixing relaxation may be enhanced
by clump-clump interactions.}


\maketitle

\section{Introduction}

In a previous work (Henriksen et al., referred to here as paper \cite{HLeDMcM09a},
and references hereafter) we discussed the %
{} relation between the formation of black holes (hereafter BH) and
of galaxies. We presented distribution functions (DF) for the co-eval
formation of spherically symmetric cusps and bulges, as formed by
radially infalling, collisionless matter (e.g. dark matter or stars).
We also discussed the influence of a centrally dominant mass, including
a point mass BH.

Observations (\cite{KR1995,Ma98,FM2000,Geb2000}) have established
a strong correlation between the BH mass and the surrounding stellar
bulge mass (or velocity dispersion), which we take to be an indication
of co-eval growth. Such growth occurs in part during the dissipative
baryon accretion by BH `seeds' in the AGN (Active Galactic Nuclei)
phase, but there is as yet no generally accepted scenario for the
origin of the seeds. Moreover, recent suggestions of very early very
supermassive BHs (e.g. \cite{Kurk2007}), together with changes in
the normalization of the BH mass-bulge mass proportionality (relatively
larger black holes at high red shift) (e.g. \cite{Mai2007}), suggest
an alternate early growth mechanism. Recently \cite{PFP2008} have
studied the possible size of the dark matter component in BH masses.
They deduce that between 1\% and 10\% of the black hole mass could
be due to dark matter.

We explored this possibility in the case of spherically symmetric
radial infall (paper \cite{HLeDMcM09a}) and in this paper we will
extend it to anisotropic infall. We use the same technique of inferring
reasonable distribution functions for collisionless matter from limits
of the time dependent Collisionless Boltzmann (CBE) and Poisson set,
and are aided by some high resolution simulations of the evolution
without a dominant central mass. We add a dominant central mass either
analytically or by iteration. Just as in the radial case, the loss
cones are not empty for substantial growth. The relaxation of collisionless
matter is due to the temporal evolution (including the radial orbit
instability) and, in addition, to possible `clump-clump' (two clump)
interactions (\cite{H2009,MH2003}).

We use, in this paper as in the previous paper (\cite{HLeDMcM09a}),
the Carter-Henriksen (\cite{CH91}) procedure to obtain a quasi-self-similar
system of coordinates (Henriksen 2006a,2006b, hereafter \cite{H2006,H2006A}).
This allows writing the CBE-Poisson set with explicit reference to
possible transient self-similar dynamical relaxation. In this way
we can remain `close' to self-similarity just as the numerical simulations
appear to do.

Studies of BH-density cusps originated with the problem of BH feeding
(\cite{P1972,BW76}), and with the notion of adiabatic growth (\cite{Y1980,Q1995,MH2002}).
Observations of the central Milky Way have detected a mainly isotropic
density cusp with logarithmic power in the range $-1.1\pm0.3$ (\cite{G2009}).
This is flatter than the adiabatic limit (\cite{MH2002}) and in any
case the adiabatic growth scenario does not produce the BH bulge correlations
(ibid). All this has spurred the investigation of co-eval dynamical
growth instead of adiabatic growth. Central cusps flatter than $-1.5$
can be created by tight binary BH systems formed in mergers that `scour'
the stellar environment (e.g. \cite{MS2006,NM99}). This process can
produce log slopes as flat as  $-1$ or even $-0.5$, and it is supported
by a strong correlation between nuclear BH mass and central luminosity
deficit (\cite{KB2009}).

However the correlation in itself only implicates the influence of
the BH. It does not necessarily require the merger history, which
in any case is unlikely to be the same for different galaxies. Consequently
we explore in this paper whether cusps as flat as those resulting
from scouring might also be produced during the dynamical formation
of the black hole.

We begin the next section with a summary of the general formulation
in spherical symmetry%
{}. Subsequently we study a system comprised of anisotropic non-radial
orbits in spherical symmetry. %
{}Finally we give our conclusions.

\section{Dynamical equations and previous results}

We will use the formulation of \cite{H2006}, wherein we transform
to infall variables the collisionless Boltzmann and Poisson equations
for a spherically symmetric anisotropic system. We begin with the
`Fujiwara' form (e.g. \cite{Fujiwara}), namely

\begin{eqnarray}
 &  & \frac{\partial f}{\partial t}+v_{r}\frac{\partial f}{\partial r}+\left(\frac{j^{2}}{r^{3}}-\frac{\partial\Phi}{\partial r}\right)\frac{\partial f}{\partial v_{r}}=0,\label{eq:Boltzmann}\\
 &  & \frac{\partial}{\partial r}\left(r^{2}\frac{\partial\Phi}{\partial r}\right)=4\pi^{2}G\int f(r,v_{r},j^{2})dv_{r}dj^{2}.\label{eq:Poisson}\end{eqnarray}

Here $f$ is the phase-space mass density, $\Phi$ is the `mean' field
gravitational potential, $j^{2}$ is the square of the specific angular
momentum and other notation is more or less standard%
{}.

The transformation to infall variables takes the form (e.g. \cite{H2006})

\begin{eqnarray}
R=r\, e^{-\alpha T/a}, &  & Y=v_{r}e^{-(1/a-1)\alpha T},\nonumber \\
Z=j^{2}e^{-(4/a-2)\alpha T}, &  & e^{\alpha T}=\alpha t,\nonumber \\
P\left(R,Y,Z;T\right)= &  & e^{(3/a-1)\alpha T}\pi f\left(r,v_{r},j^{2};t\right),\label{eq:artrans}\\
\Psi\left(R;T\right)=e^{-2(1/a-1)\alpha T}\Phi(r), &  & \Theta\left(R;T\right)=\rho(r,t)e^{-2\alpha T}.\nonumber \end{eqnarray}

The passage to the self-similar limit requires taking $\partial_{T}=0$
when acting on the transformed variables. Thus the self-similar limit
is a stationary system in these variables, which is a state that we
refer to as `self-similar virialisation' (Henriksen \& Widrow 1999,
hereafter \cite{HW99}, \cite{LeD2001}). The virial ratio $2K/|W|$
is a constant in this state (although greater than one; $K$ is kinetic
energy and $W$ is potential), but the system is not steady in physical
variables as infall continues. 

The single quantity $a$ is the constant that determines the dynamical
similarity, called the self-similar index. It is composed of two separate
reciprocal scalings, $\alpha$ in time and $\delta$ in space, in
the form $a\equiv\alpha/\delta$. As it varies it reflects all dominant
physical constants with dimensions of mass as well as of length and
time. This is because the mass reciprocal scaling $\mu$ has been
reduced to $3\delta-2\alpha$ in order to maintain Newton's constant
$G$ scale invariant (e.g. \cite{H2006}).

We assume that time, radius, velocity and density are measured in
fiducial units $r_{o}/v_{o}$,$r_{o}$, $v_{o}$ and $\rho_{o}$ respectively.
The unit of the DF is $f_{o}$ and that of the potential is $v_{o}^{2}$.
We remove constants from the transformed equations by taking \begin{equation}
f_{o}=\rho_{o}/v_{o}^{3},~~~~~~v_{o}^{2}=4\pi G\rho_{o}r_{o}^{2}.\label{eq:units}\end{equation}

These transformations convert equations (\ref{eq:Boltzmann}),(\ref{eq:Poisson})
to the respective forms

\begin{multline}
\frac{1}{\alpha}\partial_{T}P-(3/a-1)P+(\frac{Y}{\alpha}-\frac{R}{a})\partial_{R}P\\
-\left((1/a-1)Y+\frac{1}{\alpha}\left(\frac{\partial\Psi}{\partial R}-\frac{Z}{R^{3}}\right)\right)\partial_{Y}P-(4/a-2)Z\partial_{Z}P=0\label{eq:SSBoltzmann}\end{multline}
 and \begin{equation}
\frac{1}{R^{2}}\frac{d}{dR}\left(R^{2}\frac{\partial\Psi}{\partial R}\right)=\Theta.\label{eq:SSPoisson}\end{equation}
 This integro-differential system is closed by \begin{equation}
\Theta=\frac{1}{R^{2}}\int~PdY~dZ.\label{eq:dmoment}\end{equation}

This completes the formalism that we will use to obtain the results
below. The `cusps' we describe there will generally end in what is
the central `bulge' surrounding the black hole, rather than in the
black hole itself.

%
{}

\section{Spherically symmetric steady anisotropic bulges and cusps}

We follow the same pattern of discussion in this section that we used
previously (paper \cite{HLeDMcM09a}) for radial orbits%
{}. We begin in this section with steady bulge-black hole systems that
retain a memory of the prior nearly self-similar relaxation. The relaxation
proceeds by way of the relation \begin{equation}
\frac{dE}{dt}=\frac{\partial\Phi}{\partial t}|_{r}\label{eq:MVR}\end{equation}
(with the appropriate total energy and potential energy). This includes
clump-clump interactions and the radial orbit instability and produces
finally the coarse grained (i.e. steady \cite{H2006}) system.

The steady self-similar cusps were found in (Henriksen \& Widrow 1995,
hereafter \cite{HW95}) except for the case where $a=1$. However
we can recover them here by using the same procedure that we used
%
{}for radial orbits (see paper \cite{HLeDMcM09a}). We employ the characteristics
of (\ref{eq:SSBoltzmann}), together with the identity $\frac{d\Psi}{ds}=\frac{\partial\Psi}{\partial s}+\frac{dR}{ds}\partial_{R}\Psi$.
A combination of the characteristics leads to an expression for the
scaled energy on a characteristic, namely \begin{equation}
\frac{d\mathcal{E}}{ds}=-2(1/a-1)\mathcal{E}+2(1/a-1)\Psi-\frac{R}{a}\frac{\partial\Psi}{\partial R}+\frac{\partial\Psi}{\partial s},\label{eq:integralG}\end{equation}
 where \begin{equation}
\mathcal{E}\equiv\frac{Y^{2}}{2}+\frac{Z}{2R^{2}}+\Psi.\label{eq:scaleE}\end{equation}

We impose the steady state in equation (\ref{eq:integralG}) by setting
$\partial\Psi/\partial s=0$ and requiring as in paper \cite{HLeDMcM09a}
that $\Psi\propto R^{p}$ with $p=2(1-a)$ so that \begin{equation}
-\frac{R}{a}\frac{d\Psi}{dR}+2(1/a-1)\Psi=0.\label{eq:steadycond}\end{equation}
 Consequently the characteristics of equation (\ref{eq:SSBoltzmann})
are seen to imply \begin{eqnarray}
\frac{dP}{ds} & = & (3/a-1)P,\nonumber \\
\frac{d\mathcal{E}}{ds} & = & -2(1/a-1)\mathcal{E},\\
\frac{dZ}{ds} & = & -(4/a-2)Z.\nonumber \end{eqnarray}

These combine to give the general self-similar steady DF in the form
($a\ne1$ and $a\ne3/2$) \begin{equation}
P=\widetilde{P}(Z_{o},\kappa)\mathcal{E}^{q},\label{eq:genanisop}\end{equation}
 where \begin{eqnarray}
\kappa & \equiv & \mathcal{E}Z^{-\frac{1-a}{2-a}},\label{eq:defs}\\
q & \equiv & \frac{3-a}{2(a-1)}.\nonumber \end{eqnarray}
 The quantity $Z_{o}$ is the value of $Z$ at $s=\alpha T=0$, that
is $j^{2}$. However in the ab-initio steady self-similar analysis,
this dependence does not appear (\cite{HW95}). It is however present
in general in this limit from time dependent phase, and we use it
in the next paper in this series (paper \cite{HLeDMcM09c}), on non-self-similar
cusps.

The physical form of the self-similar DF becomes, on using the scaling
relations (\ref{eq:artrans}), \begin{eqnarray}
\pi f & = & \widetilde{P}(\kappa)|E|^{q},\label{eq:anisopsteady}\\
\kappa & = & |E|(j^{2})^{-(\frac{1-a}{2-a})},\nonumber \end{eqnarray}
 where $E\equiv v_{r}^{2}/2+j^{2}/(2r^{2})+\Phi$. This result has
also been shown (e.g. \cite{H2006}) to be the zeroth order DF in
the coarse graining approach that becomes the exact steady state in
the long time limit ($\alpha\rightarrow\infty$). It was also used
by Kulessa and Lynden-Bell (\cite{KL-B1992}) in their study of the
mass of the Galaxy.

An earlier important paper by Stiavelli and Bertin (\cite{StiavelliBertin})
also proposed equilibrium distribution functions for elliptical galaxies.
These were based in general on three integral models, but were reduced
to our two familiar integrals in the case of spherical symmetry. In
our case the motivation for the equilibrium DF is different, because
it derives from the self-similar dynamical growth. There are nevertheless
some significant similarities, particularly as modified to an inverted
energy distribution by Merritt, Tremaine and Johnstone (1989, hereafter
\cite{MTJ}). We discuss these points further below and in the conclusions
to this paper. 

The density profile that accompanies this DF for a particular choice
of $\widetilde{P}(\kappa)$ can be shown by direct integration over
phase space to be $\rho\propto r^{-2a}$, and subsequently the consistent
potential is $\Phi\propto r^{2(1-a)}$ so long as $a\ne1,3/2$.

The case $a=3/2$ is excluded simply because it represents a point
mass in a massless halo. However this is of interest precisely when
describing the environment of a dominant central mass, whether this
be the halo of a central black hole or the halo of a coarse-grained
collisionless bulge of stars.

The general form of the DF becomes from equation (\ref{eq:anisopsteady})
\begin{equation}
\pi f=\widetilde{P}(|E|j^{2})|E|^{3/2}.\label{eq:athreehalfs}\end{equation}
 This gives a number density of massless particles $\propto r^{-3}$
for a potential $\Phi=-M_{\star}/r$. The rms radial velocity (the
same as the radial velocity dispersion since $\overline{v_{r}}=0$)
will generally be $\propto\Phi$ for any power law choice of $\widetilde{P}(|E|j^{2})$.

One can come close to imitating the Bahcall and Wolf (\cite{BW76})
zero flux solution for a black hole cusp by choosing $\widetilde{P}(\kappa)=F(|E|j^{2})/(|E|j^{2})^{5/4}$.
This yields the DF \begin{equation}
\pi f=F(|E|j^{2})\frac{|E|^{1/4}}{(j^{2})^{5/4}},\label{eq:BW76}\end{equation}
 where $F$ is an arbitrary function. In fact, they use for argument
of their arbitrary function $\lambda=(1-e^{2})/2$, where $e$ is
the orbital eccentricity, and for Keplerian orbits around a point
mass $M_{\bullet}$ this becomes $|E|j^{2}/(GM_{\bullet})^{2}$. In
our units $GM_{\bullet}=1$.

If this were exactly the Bahcall-Wolf solution it would be remarkable,
since their solution is the result of collisional diffusion into the
loss cone. Here however we are obliged to have the additional dependence
$(j^{2})^{-5/4}$, which implies a divergent loss cone. Thus the necessary
diffusion is simply assumed in this example in order to duplicate
the energy and eccentricity dependence (arbitrary because of spherical
symmetry).

One can obtain the Bahcall-Wolf energy dependence from the general
form (\ref{eq:anisopsteady}) only by setting $a=7/3$, which is quite
unrealistic and gives the wrong arbitrary dependence. We shall see
in paper \cite{HLeDMcM09c} that it is possible to choose a simple
non-self-similar DF that does fall between these two cases and imitates
the Bahcall-Wolf result more precisely.

There are two simple limits of the form (\ref{eq:DFE}) for which
there is some evidence in the numerical simulations of isolated dark
matter bulges (\cite{MacM2006}). Just as in the radial case the following
comparison with simulations suggest that this family of DF's may actually
be realized. These limits are found by taking first $\widetilde{P}$
to be a constant $K$, and then by taking it to be proportional to
$\kappa^{-q}$. This yields respectively \begin{eqnarray}
\pi f & = & K|E|^{q},\label{eq:DFE}\\
\pi f & = & \frac{K}{(j^{2})^{w}},\label{eq:DFJ}\end{eqnarray}
 where we have defined $w=(3-a)/(4-2a)$.

We expect (see discussion to follow) that the DF (\ref{eq:DFE}) will
apply near the centre of a system where $a<1$. The self similar potential
is then increasing with radius and so must be taken positive to obtain
an attractive force. This allows us to calculate the corresponding
density explicitly as \begin{equation}
\rho=4\sqrt{2}K(\Phi)^{\frac{a}{a-1}}B(|q|-3/2,3/2).\label{eq:DFEdens}\end{equation}
 Here $B(x,y)$ is the standard Beta function, which may be expressed
in terms of the gamma function $\Gamma(x)$ as $\Gamma(x)\Gamma(y)/\Gamma(x+y)$.

The Poisson equation shows that any power law solution for the potential
with this density goes as $\Phi\propto r^{2(1-a)}$ and hence also
$\rho\propto r^{-2a}$, both as they must for self-similarity. The
only reason for giving this explicit form here is that it may be used
in an iterative calculation of the transition from halo to black hole
dominance. In such a calculation a point mass potential plus the halo
potential is inserted to give a new density, which in turn may be
used in the Poisson equation to give a new potential and so on. The
halo potential must dominate however so as to keep the net potential
positive.

As an example we assume that the potential has the form $\Phi=-C_{1}/r+C_{2}r^{2(1-a)}$
($C_{1}$ and $C_{2}$ arbitrary positive constants) so that by (\ref{eq:DFEdens})
the density becomes $\rho\propto(C_{2}-C_{1}/r^{3-2a})r^{-2a}$. From
the Poisson equation the new potential now follows as \begin{equation}
\Phi\propto\frac{C_{2}r^{2(1-a)}}{2(1-a)(3-2a)}-(C_{3}-C_{1}\ln{r})/r,\label{eq:phiiter}\end{equation}
 where $C_{3}$ should be positive to maintain the point mass contribution.
The inner boundary of this expression is at the radius where it equals
zero and the outer boundary would be inside the NFW (Navarro et al.
1996, hereafter \cite{NFW}) scale radius (which we may set equal
to our scale radius $r_{o}$). The radial mean square velocity $\overline{v_{r}^{2}}$
(also the radial squared dispersion) is proportional to $\Phi$ and
so we predict a minimum value just outside the radius $r_{m}$ where
the central mass becomes dominant (i.e., $\Phi(r_{m})=0$), followed
by a steady rise that tends to $r^{2(1-a)}$.

The density corresponding to this potential follows from (\ref{eq:DFEdens})
as \begin{equation}
\rho\propto\Phi^{\frac{a}{a-1}},\end{equation}
 so that it has for $a<1$ a rapid rise near $r_{m}$ followed by
the self-similar regime wherein $\rho\propto r^{-2a}$. The radius
$r_{m}$ would be between $1$ and $10$ pc in the halos of typical
galaxies.

Of course this iteration has not fully converged but subsequent cycles
will produce higher order terms. The result (\ref{eq:phiiter}) seems
to indicate that the black hole mass is enhanced by an $r^{-3}$ cusp
that peaks just outside $r_{m}$. 

\begin{figure}[t]
\includegraphics[width=1\columnwidth]{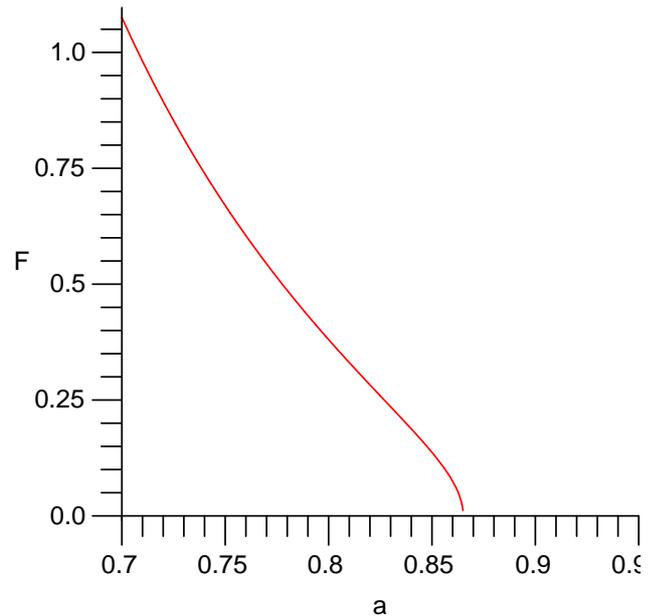}

\caption{\label{fig:Fa} The curve shows the dependence of the factor that
multiplies $\Phi$ in the tangential velocity dispersion as a function
of $a$. The adiabatic variation of $a$ with $r$ is approximately
$a=r^{0.2}$ (\cite{H2007}), where $r$ is measured in units of the
NFW scale radius $r_{s}$.}

\end{figure}

The radial velocity dispersion is generally equal to the root mean
square radial velocity for the DF (\ref{eq:DFE}) since the mean radial
velocity is zero. This is proportional to $\Phi\propto r^{2(1-a)}$.
The squared tangential velocity dispersion however becomes \begin{equation}
\sigma_{\perp}^{2}=\Phi\left(\frac{8~B(|q|-5/2,5/2)}{3~B(|q|-3/2.3/2)}-\frac{\pi^{2}B(|q|-2,2)^{2}}{8~B(|q|-3/2,3/2)^{2}}\right),\label{eq:DFEtdispersion}\end{equation}
 where $q<-2$ or $1>a>1/3$. The factor $F(a)$ in this equation
that multiplies $\sqrt{\Phi}$ is shown in figure (\ref{fig:Fa})
as a function of $a$.

 Unlike the radial velocity dispersion, which should steadily increase
with \emph{radius} proportionally to $\Phi$ (which flattens as $a\rightarrow1$),
the tangential velocity dispersion is expected to show a more precipitous
drop as $r$ tends towards the scale radius and $a\rightarrow1$ as
$F(a)$ declines. The variation of $a$ is only adiabatic with radius
however, varying roughly as $r^{0.2}$. We shall see evidence for
this below.

We turn first to an examination of the other limit (\ref{eq:DFJ}).
We expect this to apply in the outer region to which much angular
momentum has been transferred by the bar due to the radial orbit instability
(MacMillan et al. 2006, hereafter \cite{MWH2006}). Hence we choose
the regime where $a>1$ and the energy/potential is negative. Iteration
is not really relevant in this limit since we are far from the black
hole, but because of its importance in calculating mean quantities
we give the density as \begin{equation}
\rho=\frac{2^{(3/2-w)}}{3/2-w}KI_{0}(w)r^{-w}|\Phi|^{(3/2-w)}.\end{equation}
 Once again the Poisson equation has the consistent solution $\Phi\propto-r^{2(1-a)}$
and $\rho\propto r^{-2a}$. The integral $I_{0}(w)$ is given by \begin{equation}
I_{n}(w)=\int_{0+}^{1}~\frac{dy}{y^{w}}(1-y)^{(n-1/2)}\end{equation}
 where the lower limit must be greater than zero for convergence since
$w\ge1$ for $a\ge1$. 

The mean square radial velocity becomes in this limit \begin{equation}
\overline{v_{r}^{2}}=\frac{2(3/2-w)}{5/2-w}\frac{I_{1}}{I_{0}}|\Phi|,\label{eq:DFJrdispersion}\end{equation}
 where, since $a>1$, this decreases with radius. The tangential velocity
dispersion has once again a more complicated expression namely \begin{multline}
<\sigma_{\perp}^{2}>=|\Phi|\left(\frac{2(3/2-w)}{5/2-w}\frac{I_{0}(w-1)}{I_{0}(w)}\vphantom{\left(\frac{2^{5/2}I_{0}}{I_{0}}\right)^{2}}\right.\\
\left.-\left(\frac{2^{5/2}(3/2-w)I_{0}(w-1/2)}{(2-w)I_{0}(w)}\right)^{2}\right).\label{eq:DFJtdispersion}\end{multline}
 Although $w(a)$ is slowly varying for $1<a<3/2$, the factor multiplying
$\Phi$ declines to zero as $w\rightarrow3/2$. 

It is time however to compare these DF's chosen for their self-similar
memory to the results of simulations, in order to justify our attention.

In references (\cite{H2006}) and (Henriksen 2007, hereafter \cite{H2007})
it was concluded that $a\approx0.72$ near the outer boundary of the
relaxed region. Moreover we know from those papers, and more generally
from extensive cosmological simulations, that $a\approx0.5$ in the
interior, well relaxed, region. This is referred to as adiabatic self-similarity
(\cite{H2006}) since the index $a$ dynamically evolves, but relatively
slowly ($a\propto r^{\alpha_{F}},~\alpha_{F}=O(0.2)$, \cite{H2007})
as already employed above. To match the simulation results we require
$a\approx3/2$ in the vicinity of the NFW scale radius, after which
there may be a tidal truncation to $r^{-4}$ (Henriksen 2004, hereafter
\cite{H2004})

There is evidence in the simulations for the persistence of self-similarity,
even in the case that most closely resembles an isolated cosmological
halo (i.e. cosmological perturbations are included in the initial
conditions). Thus in the cosmological-like halo simulation of (\cite{MacM2006}),
the virial mass and virial radius are found to grow in a power law
manner according to $t^{2.16}$ and $t^{1.30}$ (we do not quote numerical
errors, which are at the level of a few percent) respectively. The
logarithmic density slope in the outer relaxed region is approximately
$-1.4$ while the pseudo-density logarithmic slope is $-2.16$.%
{}

The numerical predictions follow by using $a=0.72$ from (\cite{H2007}).
The self-similar mass growth inside any growing radius (fixed $R$)
is (\cite{HW99}) $\propto t^{(3/a-2)}\propto t^{2.16}$, while $r\propto t^{1/a}\propto t^{1.38}$
(see equation (\ref{eq:artrans})). The logarithmic density slope
is predicted to be $-2a=1.44$, and the pseudo-density logarithmic
power (e.g \cite{H2007}, \cite{Hansen2004}) is given by $a-3=-2.28$.
These reasonable correspondences with the numerical simulation (only
the predicted pseudo-density differs significantly from the simulated
value) encourage us to adopt a DF as one of the above steady forms.
The self-similar memory is incorporated either in $a=0.72$ or in
$a=0.5$ in the relaxed region, which boundary is approximately (in
fact the region where $a<1$ extends beyond the scale radius by about
a factor $2$ in the simulation) coincident with the NFW scale radius.

\begin{figure}[t]
\includegraphics[width=1\columnwidth]{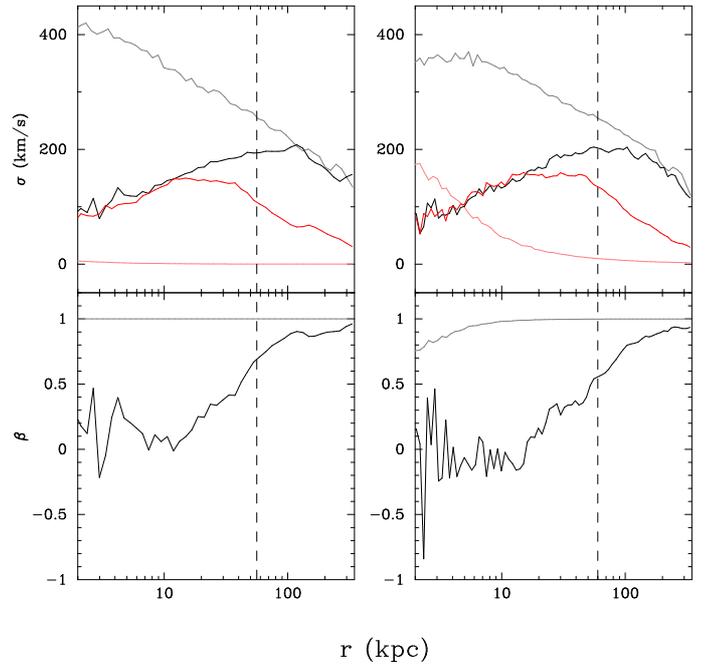}

\caption{\label{fig:sigma} The figure shows two distinct simulations from
(\cite{MacM2006}). The right hand panels display the result (with
continuing infall) of an isolated halo simulation starting from a
set of particles perturbed by a cosmological spectrum of density and
velocity fluctuations. The panel on the left indicates the same stage
as developed from a non-perturbed halo. The upper line on the upper
panel in each case is for a similar halo, but allowing only purely
radial collapse. The middle line is the radial velocity dispersion
while the lower line is the tangential dispersion. The lower panels
display the anisotropy parameter for non-radial collapse.}

\end{figure}

The velocity dispersion in the relaxed region is based on the DF (\ref{eq:DFE})
and has been given in equation (\ref{eq:DFEtdispersion}) for the
tangential case, while the radial dispersion is proportional to $\sqrt{\Phi}\propto r^{1-a}$.
Hence, away from the transition to the black hole region, we can test
our predicted dispersions against the simulation . Using the value
$a=0.72$ this gives the radial ($\sigma_{r}$) velocity dispersion
going as $r^{0.28}$ which should be compared with figure (\ref{fig:sigma}).
The cosmological case is the right hand panel, which the phase space
diagram (not shown) shows to be more relaxed than the initially unperturbed
simulation on the left. The $r^{0.28}$ rise is a reasonable description
of the numerical behaviour, especially as some adiabatic evolution
in $a$ towards unity is to be expected. This will flatten the rate
of rise Between $4$ and $60$ kpc, the radial dispersion rises by
about a factor $2$ while the predicted value using $a=0.72$ is $2.13$.
Allowing $a$ to increase only to $0.75$ on average, improves the
fit.

Eventually at large enough radius, according to the complete cosmological
simulations, we require $a\rightarrow3/2$ and so we predict a decline
in the radial dispersion proportional to $r^{-1/2}$ in this region.
Figure (\ref{fig:sigma}) indicates that this begins at about twice
the indicated scale radius. Before this point but still outside the
scale radius the decline is slower. In fact the simulation gives $a\approx5/4$
in this region so that the expected decline would be only as $r^{-1/4}$.

The tangential dispersion velocity has the same behaviour at small
$r$ but rolls over more rapidly. This is indicated very clearly in
the graphs of the anisotropy parameter $\beta\equiv1-\sigma_{\perp}^{2}/\sigma_{r}^{2}$
on the panel below the velocity dispersion. This is consistent qualitatively
with the action of the factor $F(a)$ defined in equation (\ref{eq:DFEtdispersion}),
and displayed in figure (\ref{fig:Fa}) over the relevant range of
$a$. The weak adiabatic dependence of $a$ on $r$ is not weak enough
however to cover the range in $r$ over which $\sigma_{\perp}$ stays
flat. There is some uncertainty in this factor and for $a\propto r^{0.15}$
one arrives at nearly a factor two in radius as $a$ varies from $0.72$
to $0.8$. However figure (\ref{fig:Fa}) shows that $F$ has dropped
by more than a factor two in this range, while $\sqrt{\Phi}$ increases
by only a factor of $1.2$ at best. The implied factor two decline
is not seen. The only explanation is that $a$ does not vary significantly
in this region, so that there is little adiabatic evolution.

At the end of this plateau we have entered the region where $a\approx5/4>1$.
The decline in $\sqrt{\Phi}$ is however only as $r^{-0.25}$ there,
whereas the observed decline in figure (\ref{fig:sigma}) is more
like $r^{-1}$. Once again the difference between this behaviour and
that of the radial dispersion must lie in the factor multiplying $\Phi$.%

So we have found above that the implications of equations (\ref{eq:DFE})
and (\ref{eq:DFJ}) are reasonable, but do we have any evidence concerning
the DF itself?

\begin{figure}[t]
\includegraphics[width=1\columnwidth]{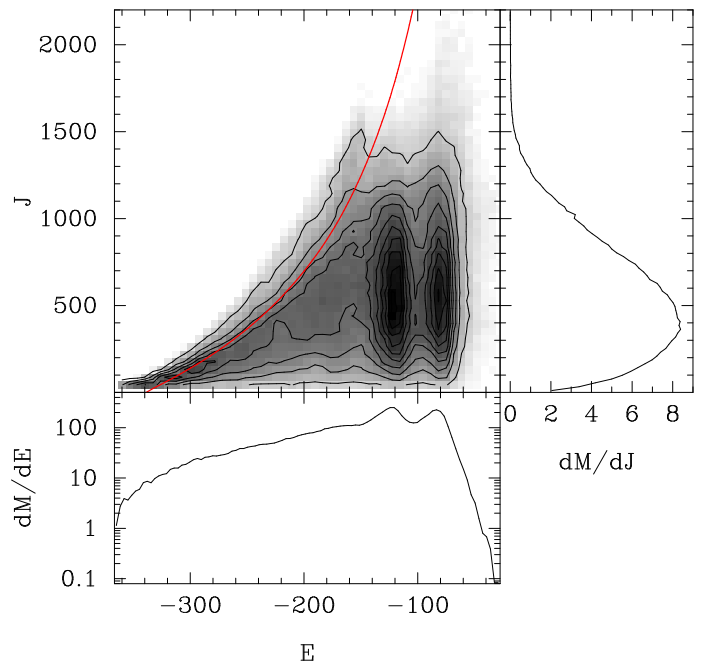}

\caption{\label{fig:dMdE} The figure shows a contour plot of $d^{2}M/dEdJ$
for the full simulation of an isolated halo (\cite{MacM2006}). The
separate plots of $dM/dE$ and $dMdJ$ are also indicated. The corresponding
density of states is almost independent of $J$ and is the expected
(e.g.\cite{BT1987}) $|E|^{-2.5}$ law. The line on the upper left
is $J(E)$ for circular orbits.}

\end{figure}

In the outer part of the relaxed region MacMillan (ibid, and figure
\ref{fig:dMdE}) finds that the DF is mainly dependent on angular
momentum. This is evident from the figure by noting that numerically
$g(E,j^{2})\propto dM/dE$%
{} in the range $-100$ to $-200$ energy units (\cite{MacM2006}).
Hence from the definition\begin{equation}
f(E,j^{2})=\frac{1}{g(E,j^{2})}\frac{d^{2}M}{dEdj^{2}},\end{equation}
we infer that $f\propto d\ln{dM/dE}/dj^{2}$. This is independent
of energy in the outer region. Note that in terms of $j^{2}$ the
figure is simply stretched by a factor of $2$ in the $j$ direction. 

We therefore fit the self-similar DF (\ref{eq:DFJ}) with $a=1.25$
to obtain $f\propto(j^{2})^{-1.6}$. This is in fact a reasonable
fit to the DF in the less tightly bound region with angular momentum
greater than about $500$ units (figure \ref{fig:dMdE}).%
{}\textbf{ }Such a cut-off in angular momentum was contained in the
Stiavelli and Bertin DF (\cite{StiavelliBertin}) where it was exponential.
\cite{MTJ} gave a heuristic justification for a cut-off in $j^{2}$
and also wondered whether an exponential or a power-law cut-off was
superior. They also asked whether such behaviour arose naturally in
the course of the dynamical formation of galaxies. Here our answer
is in favour of a power-law cut-off following self-similar infall. 

An intermediate region where the logarithmic density slope varies
from just above $2$ to about $3$ does occur in all simulations of
collisionless matter. Moreover this region tends to coincide with
the passage from isotropy to radial orbits (e.g. \ref{fig:sigma}).
According to the preceding remarks (i.e. the DF (\ref{eq:DFJ}) with
$a>1$) this region can be characterized by roughly constant numbers
of particles over a broad range of energy, but with numbers rapidly
increasing towards a minimum in angular momentum. Some particles would
have too much angular momentum to enter the central region, while
those below a critical angular momentum can populate the central regions.

In the simulations this redistribution of angular momentum is attributed
to the onset of the Radial Orbit Instability (ROI) in (\cite{MWH2006})
and the consequent formation of a bar. In the current spherically
symmetric analysis, we are forced to patch these different regions
together `by hand', guided only by the different possibilities for
$a$.

In the inner relaxed region we expect the DF (\ref{eq:DFE}) to apply
with $a\approx0.72$. This gives an isotropic $f\propto E^{-4}$.
The simulations (\cite{NFW}) suggest that $a\rightarrow1/2$ in the
very centre of the system, which would give $f\propto E^{-5/2}$.
This is in accord with previous discussions (e.g. \cite{H2006}).
The fact that our energy is positive, while the simulation energy
is negative, reflects the different zero points for the potential.
We use the centre of the system as the reference zero in our analysis,
while the simulation uses infinity.

Numerically, $g(E)\propto|E|^{-2.5}$ in this region so that $d^{2}M/dEdj^{2}$
should vary as $E^{-6.5}$ ($E^{-5}$ if $a=0.5$). There is very
little dependence on $j^{2}$ here so this prediction might be compared
to the behaviour of $dM/dE$. This quantity is falling steeply beyond
$E=-300$ (\ref{fig:dMdE}), but the evidence is weak at the limit
of resolution. 

At slightly higher angular momentum there is a rising dependence on
$j$. This can be imitated with a self-similar solution by taking
$\widetilde{P}(\kappa)\propto1/\kappa$ in equation (\ref{eq:anisopsteady}),
which gives the DF $f\propto E^{-5}(j^{2})^{1/5}$.

The above prolonged discussion suggests that a DF with a self-similar
memory can be used to parametrise the phase space of these collisionless
simulations. At least this is true if we accept the adiabatic variability
of $a$, which has been made plausible on the grounds of a local entropy
maximum (e.g. \cite{H2007}). However what has this to do with the
black-hole bulge mass correlation that the co-eval growth was supposed
to explain? 

If one assumes that the halo around the seed black hole is accreted
eventually through two-body and clump-clump relaxation (\cite{MH2003}),
then the mass of the black hole/halo should be proportional to the
mass of the bulge simply because of the power law density (with nearly
constant power inside $r_{s}$). Letting $M_{\bullet}$ be the black
hole mass, $M_{s}$ be the bulge mass and $r_{h}$, $r_{s}$, be the
black hole halo and bulge radii respectively ($r_{s}$ is the scale
radius); we find the relation \begin{equation}
\frac{M_{\bullet}}{M_{s}}=\left(\frac{r_{h}}{r_{s}}\right)^{(3-2a)}.\label{eq:bhmass}\end{equation}
 If the appropriate $a=0.72$, then the power of the radial dependence
is $1.56$. For a true universal relation however, we must assume
that galaxy growth is not only self-similar, but that there is a similarity
between galaxies so that the scale ratio is similar. This is not exactly
the case (e.g. \cite{diverse2009}), but it is nearly so. Such a relation
might be tested by high resolution simulations containing a seed black
hole.

\subsection{More general self-similar distribution functions}

The limits of equation (\ref{eq:anisopsteady}) that we discussed
above are quite arbitrary, except for the expectation of central isotropy
and the dominance of angular momentum in the outer regions. In this
section we explore more general possibilities, even though empirically
the preceding limits seem adequate. 

The distribution function (\ref{eq:anisopsteady}) can be put in a
form that seems to generalize the FPDF (i.e. Fridmann and Polyachenko
DF%
{}). We choose $\widetilde{P}\propto\kappa^{-\left(\frac{1}{a-1}\right)}$
to obtain \begin{equation}
\pi f=\frac{K}{(j^{2})^{\frac{1}{2-a}}|E_{o}-E|^{1/2}}.\label{eq:genFPDF}\end{equation}

The density integral over this DF requires a lower cut-off in angular
momentum for $a>1$ and an upper cut-off in energy for $a<1$ in order
to avoid singularities. We have exhibited the upper energy cut-off
in (\ref{eq:genFPDF}) explicitly. It appears as the arbitrary constant
in the potential so that the difference $E_{o}-E$ still has the self-similar
behaviour ($r^{2(1-a)}$), as does the density ($r^{-2a}$). Moreover
for $a>1$ the minimum angular momentum should be a fixed fraction
of the value $2r^{2}(E-\Phi)$, say $0<k_{1}\ll1$, in order for the
density profile and potential associated with equation (\ref{eq:anisopsteady})
to apply. The radial velocity dispersion is also $\propto r^{2(1-a)}$.
It is only when $a=1$ that one obtains the $r^{-2}$ profile of the
radial FPDF, and this must be discussed separately below.

If one computes formally the density by integrating the DF (\ref{eq:genFPDF})
in a negative energy region (negative although $a<1$, because the
central mass dominates), one finds that for $a<1$ \begin{equation}
\rho\propto r^{-\frac{2}{2-a}}(|\Phi|-|E_{o}|)^{\frac{1-a}{2-a}}.\label{eq:zerodens}\end{equation}
As in the previous section, this can be used in an iterative fashion.

Thus in a region dominated by a central point mass (and taking $E_{o}=0$)
we see that the cusp would be, at lowest order, \begin{equation}
\rho\propto r^{-\frac{3-a}{2-a}}.\end{equation}
 The next order would be found by using this in the Poisson equation
to obtain a new potential and then a new density from equation (\ref{eq:zerodens}).
Once again this low order result can not be flatter than $r^{-1.5}$.The
steepest limit is $r^{-2}$ as $a\rightarrow1-$. We remark that at
$a=2/3$ one obtains the Bahcall and Wolf zero flux cusp $r^{-7/4}$,
due to the filled loss cone in the DF. For $a=0.72$, the cusp is
like $r^{-1.78}$, very close to the Bahcall and Wolf cusp and still
with a filled loss cone. In paper \cite{HLeDMcM09c} of this series
we shall find an exact example of this type.

\subsection{Singular isothermal sphere: the global inverse square law}

So far we have ignored the case when $a=1$, the singular isothermal
limit. In fact the steady DF with $a=1$ must be treated separately
because of the logarithmic potential $\Phi=\Psi\ln{r}$ ($\Psi$ constant)
that accompanies the inverse square density law. We may still deduce
it from the characteristics of (\ref{eq:SSBoltzmann}) if we take
$T=0$ finally, since a steady state is the same at any time. The
appropriate characteristics become, with $a=1$, \begin{eqnarray}
\frac{dP}{ds} & = & 2P,\nonumber \\
\frac{dR}{ds} & = & \frac{Y}{\alpha}-R,\label{eq:anchars}\\
\frac{dY}{ds} & = & \frac{1}{\alpha}\left(\frac{Z}{R^{3}}-\frac{\Psi}{R}\right),\nonumber \\
\frac{dZ}{ds} & = & -2Z.\nonumber \end{eqnarray}

These integrate to give \begin{eqnarray}
P & = & \widetilde{P}(Z_{o},\kappa)/Z,\\
\kappa & = & \mathcal{E}-\frac{\Psi}{2}\ln{Z},\nonumber \end{eqnarray}
 where \begin{equation}
\mathcal{E}\equiv\frac{Y^{2}}{2}+\frac{Z}{2R^{2}}+\Psi\ln{R},\end{equation}
 and $Z_{o}$ is again $j^{2}$ at $s=0$ on the characteristic.

By writing the self-similar physical form at $T=0$ we find (for $a=1$)
\begin{eqnarray}
\pi f & = & \frac{\widetilde{P}(\kappa)}{j^{2}},\label{eq:anisopsteadya}\\
\kappa & = & \frac{v_{r}^{2}}{2}+\frac{j^{2}}{2r^{2}}+\Psi\ln{r}-\frac{\Psi}{2}\ln{j^{2}}.\nonumber \\
 & \equiv & \frac{v_{r}^{2}}{2}+\frac{j^{2}}{2r^{2}}-\frac{\Psi}{2}\ln{\frac{j^{2}}{r^{2}}}.\end{eqnarray}
 Once again $E\equiv v_{r}^{2}/2+j^{2}/(2r^{2})+\Psi\ln{r}$ so that
$\kappa=E-\Psi/2\ln{j^{2}}$.

This limit with $a=1$ is an extension of the inverse square density
law to non-radial orbits, but the function of $\kappa$ is arbitrary.
The DF is certainly non-unique for the same density profile. 

If we calculate the density profile for $\kappa>0$ by integrating
over (\ref{eq:anisopsteadya}) we arrive eventually at \begin{equation}
\rho=\frac{\sqrt{2}}{r^{2}}\int_{0}^{\infty}~~\frac{\widetilde{P}(\kappa)}{\kappa^{1/2}}~d\kappa\int_{u_{1}}^{u_{2}}~\frac{du}{u\sqrt{1-(u-\Psi\ln{u})/2\kappa}},\end{equation}
 where $u\equiv j^{2}/r^{2}$. The limits $u_{1}$, $u_{2}$ are the
two roots found in terms of the Lambert function by setting the argument
of the square root equal to zero. When $\kappa/\Psi$ is large, the
lower root becomes zero and the upper root approaches $2\kappa$.
There is slow convergence of the inner integral. Provided that $\widetilde{P}(\kappa)$
is such as to make the outer integral converge, the profile is $r^{-2}$.

We note that for any finite $\kappa$, the possible range of $u$
has a definite upper limit and a definite lower limit that may be
close to zero. This means that at given $r$ there is a finite range
in $j^{2}$ that is present, and that this range collapses on zero
as $r\rightarrow0$. This is as expected with spherical symmetry.

It is possible to have solutions with $\kappa<0$ whereupon \begin{equation}
\rho=\frac{\sqrt{2}}{r^{2}}\int_{0}^{(\Psi\ln{\Psi}-\Psi)/2}~~\frac{\widetilde{P}(\kappa)~d|\kappa|}{|\kappa|^{1/2}}~~\int_{u_{1}}^{u_{2}}~~\frac{du}{u\sqrt{(\Psi\ln{u}-u)/2|\kappa|-1}},\end{equation}
 where again the inner limits are the roots found by setting the argument
of the square root equal to zero. For a real range of values, $\Psi/|\kappa|$
must be greater than a minimum value that depends on the value of
$\Psi$. The same restriction on angular momentum as $r\rightarrow0$
applies. In all cases the velocity dispersion is constant. 

Unlike the radial FPDF that has an inverse square cusp, a black hole
is not readily incorporated into this distribution. However at large
enough $r$ (or basically where in phase space where $v_{r}/r^{2}\ll v_{o}/r_{o}^{2}$)
it suffices to modify the integral by a point mass potential so that
\begin{equation}
\kappa=\frac{v_{r}^{2}}{2}+\frac{j^{2}}{2r^{2}}+\Psi\ln{r}-\frac{GM_{\bullet}}{r}-\frac{\Psi}{2}\ln{j^{2}}.\label{eq:isobh}\end{equation}
 The velocity dispersion is no longer strictly constant, unless $j^{2}\gg GM_{\bullet}r$,
but the influence of the black hole is weak in this regime.

One special model of an inverse square, steady, anisotropic bulge
is provided by equation (\ref{eq:anisopsteadya}) by taking \begin{equation}
\widetilde{P}(\kappa)=K\exp{\left(-(\frac{2\kappa}{\Psi}(1+b))\right)},\label{isoGauss}\end{equation}
 where $b$ is any real number.

In velocity space the anisotropic DF becomes explicitly ($v^{2}\equiv v_{r}^{2}+v_{\perp}^{2}$)
\begin{equation}
\pi f=\frac{K}{r^{2}}\exp{\left(-(\frac{(b+1)v^{2}}{\Psi})\right)}(v_{\perp}^{2})^{b}.\label{newGauss}\end{equation}
This form is quite closely related to the DF suggested by Stiavelli
and Bertin (\cite{StiavelliBertin}). However it offers a combination
of an exponential cut-off and a possible power-law cut-off in angular
momentum. It does have the inverted distribution in energy (it increases
outward) as suggested by \cite{MTJ}. This is also true of the DF
that we suggest for the inner region, namely equation (\ref{eq:DFE})
with $a<1$. Thus we obtain this inversion naturally as a consequence
of the self-similar evolution. 

When $b=0$ we have a pure Gaussian in the energy and once again the
singular isothermal sphere. If $b<0$ there is a filled loss cone
and for $b>0$ the loss cone is empty. Provided $b>-1$ this DF is
elliptical in velocity space ($v_{r}$, $v_{\perp}$ ) at each $r$
and becomes circular (isotropic) at $b=0$. The isothermal form with
$b=0$ also appears in the fully asymmetric case of zeroth order coarse
graining as discussed in the appendix of (\cite{H2004}).

This example is only of interest in the context of the persistent
$r^{-2}$ mass distribution in galaxies and clusters referred to in
paper \cite{HLeDMcM09a}. There is clearly a lack of uniqueness however,
as several distribution functions may produce an inverse square density
profile. Adding a black hole to this example is inconvenient (unless
it is negligible) as the form of the DF is not suitable for iteration.
However just as in the general case above, $\kappa$ may be taken
as that given in equation (\ref{eq:isobh}).

{}

One can calculate the growth of a central black hole embedded in an
inner bulge with the DF (\ref{eq:DFE}). We can expect to maintain
the steady bulge DF until the black hole mass is a significant fraction
of the bulge mass. %
{}We find after obvious but tedious calculations that \begin{equation}
\frac{dx}{dt}=(2\pi)^{3/2}\sqrt{G\rho_{s}}\frac{6G\rho_{s}r_{s}^{2}}{c^{2}}x^{3(1-a)},\end{equation}
 where $x\equiv r_{\bullet}/r_{s}$, the subscript $s$ refers to
characteristic bulge quantities, and $a>1/3$ for convergence of the
integrals. We recall that the last stable radius of a Schwarzschild
black hole is $r_{\bullet}\equiv6GM_{\bullet}/c^{2}$. For $a=2/3$
the growth is exponential, but the e-folding time is longer than the
age of the Universe ($5\times10^{19}s$) even if the bulge has $10^{10}M_{\odot}$
in one kiloparsec. The result is similar for other values of $a$.%
{} However growth of halo cloaking the actual black hole could be much
faster. In general the mass growth from the DF (\ref{eq:DFE}) is
($r_{h}$ and $M_{h}$ are replaced by the black hole values to obtain
the previous equation) \begin{equation}
\frac{dM_{h}}{dt}=(2\pi)^{3/2}\sqrt{G\rho_{s}}\rho_{s}r_{s}^{3}\left(\frac{r_{h}}{r_{s}}\right)^{3(1-a)},\end{equation}
 where $M_{h}$ is the mass accreted inside the radius $r_{h}$. This
radius would be an intermediate scale between the bulge scale $r_{s}$
and $r_{\bullet}$, determined ultimately by the minimum angular momentum
actually available. The accretion rate is proportional to the radius
of the structure and is faster than the black hole rate by $r_{h}/r_{\bullet}$
when $a=1$. One assumes that all of this mass is eventually accreted
by the black hole through relaxation processes (e.g. \cite{MH2003}),
although multiple steps might be required.Then the black hole growth
is limited by the time to empty this reservoir onto the centre.

This calculation is done without any bias towards the loss cone ($j\approx0$),
and with a self-similar DF of the form (\ref{eq:genFPDF}), one can
actually grow the black hole directly. We shall save this calculation
for paper \cite{HLeDMcM09c}.

\section{Conclusions}

In this paper, we have developed distribution functions that describe
both dark matter bulges and a central black hole or at least a central
mass concentration. We succeed mainly in describing the dark matter
bulges.

In the discussion of cusps and bulges based on purely radial orbits
(paper \cite{HLeDMcM09a}), we were able to distinguish the Distribution
function of Fridmann and Polyachenko %
{} from that of Henriksen and Widrow%
{}. The FPDF was found to describe accurately the purely radial simulations
of isolated collisionless halos carried out in (\cite{MacM2006}).
Moreover a point mass could be added without changing the form of
the DF, which allows a self-similar growth of a central bulge or black
hole. 

The final result concerning steady, self-similar radial orbits concerned
the special case $a=1$. The DF is a Gaussian that has been found
previously in coarse graining (\cite{HLeD2002}). We included it there
as a second example of a radial DF that produces an $r^{-2}$ density
profile (\cite{Mutka09}).

The inclusion of angular momentum here leads to more realistic situations.
We re-derived the steady self-similar DFs from first principles in
equation (\ref{eq:anisopsteady}). We showed that these can be used
to describe the simulated collisionless halos calculated in (\cite{MacM2006}),
if we use the value of $a\approx0.72$ identified in (\cite{H2007})
and take two limits. In one (\ref{eq:DFE}) the DF is isotropic and
describes approximately the most relaxed central region of the bulge.
The other limit (\ref{eq:DFJ}) describes the outer region and the
transition across the NFW scale radius. The qualitative correspondence
supports the relevancy of this family. The potential of a central
mass can not be included exactly in these distribution functions,
but iteration is possible.%
{} An example was given for the DF (\ref{eq:DFE}).

In particular, velocity dispersions have been calculated and compared
qualitatively with the results of a high resolution numerical simulation
of an isolated halo. These correspond to reasonable densities when
adiabatic self-similarity is employed. Unfortunately these do not
apply to the vicinity of the black hole itself. However the eventual
dominance of the black hole through the $r^{-1}$ potential becomes
obvious, as seen in the iterated potential (\ref{eq:phiiter}).

Simple forms for the DF have been suggested for the various regions
of the halo. And given the dynamic co-eval growth of black-hole/halo
and bulge, we explain the black-hole bulge mass correlation simply
(\ref{eq:bhmass}). This relies on the approximate similarity between
different halos.

The forms that we find follow from the assumed self-similar path to
equilibrium. Remarkably they are in accord with the general forms
for the DF as presented in Stiavelli and Bertin (\cite{StiavelliBertin})
and as modified in \cite{MTJ}. That is, these distribution functions
contain an inner inverted distribution in energy (referred to as negative
temperature in \cite{MTJ}) and a cut-off in the distribution in angular
momentum farther from the centre. The scale of the inner region that
is argued to coincide with NFW scale radius in this paper is the parameter
$r_{a}$ in the earlier papers, so inner and outer have similar meanings. 

The variation is a power-law generally in both energy and angular
momentum, although in the very relevant case of an inverse square
density profile, an exponential variation is possible. These conclusions
follow from the assumed adiabatically self-similar evolution towards
equilibrium. As such they tend to answer questions posed in \cite{MTJ}
such as whether these distributions arise naturally and if the variations
are exponential or power-law. We now know however that part of the
answer lies in the anisotropy produced by the radial orbit instability
(\cite{MWH2006}). 

We have also found a self-similar generalization of the radial FPDF
in equation (\ref{eq:genFPDF}). Unfortunately this DF does not have
the property of yielding a density that is independent of the potential
and so a point mass potential is incompatible with self-similarity.
It might describe a system at large radii with a large central mass.

As always $a=1$ must be treated separately and we give a derivation
from first principles. The result is new and it generalizes the FPDF
in the sense that $\rho\propto r^{-2}$ always. Although we can not
place a central mass inside this bulge exactly, it allows an anisotropic
DF in the $r^{-2}$ bulge region.

Generally we find that we can not describe elegantly anisotropic bulges
containing black holes with self-similar DFs. %
{}

In the next paper in this series (paper \cite{HLeDMcM09c}), we shall
widen our scope to the study and production of non-self-similar cusps
and DF.

\section{Acknowledgements}

RNH acknowledges the support of an operating grant from the canadian
Natural Sciences and Research Council. The work of MLeD is supported
by CSIC (Spain) under the contract JAEDoc072, with partial support
from CICYT project FPA2006-05807, at the IFT, Universidad Autonoma
de Madrid, Spain

\end{comment}
{}

\bibitem[Ferrase \& Merritt 2000]{FM2000}Ferrarese,L.,\& Merritt,
D., 2000, ApJ,539, L9.


\bibitem[Fridmann \& Polyachenko 1984]{FP1984}Fridman,A.M., \& Polyachenko,
V.L., 1984,\textit{Physics of Gravitating Systems}, Springer, New
York.

\bibitem[Fujiwara 1983]{Fujiwara}Fujiwara, T., 1983, PASJ, 35, 547.

\bibitem[Gebhardt et al. 2000]{Geb2000}Gebhardt,K., et al.,2000,
ApJ,539,L13.

\bibitem[Gillessen et al. 2009]{G2009} Gillessen, S., et al., 2009,
ApJ, 692, 1075.

\bibitem[Hansen 2004]{Hansen2004} Hansen, S., 2004, MNRAS, 352, L41.

\bibitem[HW95]{HW95}Henriksen, R.N., Widrow, L.M., 1995, MNRAS, 276,
679.%
{}

\bibitem[HW 1999]{HW99}Henriksen, R.N., Widrow, L.M., 1999, MNRAS,
302, 321.

\bibitem[Henriksen \& Le Delliou 2002]{HLeD2002}Henriksen, R.N.,
\& Le Delliou, M., 2002, MNRAS, 331, 423.

\bibitem[H2004]{H2004} Henriksen, R.N., 2004, MNRAS, 355, 1217.

\bibitem[H2006a]{H2006}Henriksen, R.N., 2006, ApJ, 653,894.

\bibitem[H2006b]{H2006A} Henriksen, R.N.,2006, MNRAS, 366, 697.

\bibitem[H2007]{H2007} Henriksen, R.N., 2007, ApJ,671,1147.

\bibitem[Henriksen 2009]{H2009} Henriksen, R.N., 2009, ApJ, 690,
102.

\bibitem[Kulessa \& Lynden-Bell 1992]{KL-B1992} Kulessa, A.S.\& Lynden-Bell,
D., 1992. MNRAS,255,105. 

\bibitem[Kurk et al. 2007]{Kurk2007} Kurk, J.D., et al., 2007, ApJ,
669, 32.


\bibitem[Kormendy \& Richstone 1995]{KR1995}Kormendy, J.,\& Richstone,
D., 1995, Ann.Rev.A\&A.

\bibitem[Kormendy \& Bender 2009]{KB2009}Kormendy, J., \& Bender,
R., 2009, ApJ, 691,L142.


\bibitem[Le Delliou 2001]{LeD2001}Le Delliou, M., 2001, PhD Thesis,
Queen's University, Kingston, Canada.

\bibitem[I]{HLeDMcM09a}Le Delliou, M., Henriksen, R.N., \& MacMillan,
J.D., 2010, arXiv: 0911.2232 (I)%
{}

\bibitem[III]{HLeDMcM09c}Le Delliou, M., Henriksen, R.N., \& MacMillan,
J.D., 2009, submitted {[}arXiv : 0911.2238{]} (III)%
{}

\bibitem[MacMillan \& Henriksen 2002]{MH2002} MacMillan, J.D., \&
Henriksen, R.N., 2002, ApJ, 569,83.

\bibitem[MacMillan \& Henriksen 2003]{MH2003} MacMillan, J.D., \&
Henriksen, R.N., 2003, Carnegie Observatories Astrophysics Series,
1, \textit{Co-evolution of Black Holes and Galaxies}, L.C. Ho

\bibitem[MacMillan 2006]{MacM2006}MacMillan, J., 2006, PhD Thesis,
Queen's University at Kingston, ONK7L 3N6, Canada.

\bibitem[MWH 2006]{MWH2006} MacMillan, J.D., Widrow, L.M., \& Henriksen,
R.N., 2006, ApJ,653, 43.

\bibitem[Magorrian et al. 1998]{Ma98}Magorrian, J., et al., 1998,
AJ, 115, 2285.

\bibitem[Maiolino et al. 2007]{Mai2007}Maiolino, R., et al.,2007,
A\&A, 472, L33.

\bibitem[Merritt \& Szell 2006]{MS2006} Merritt, D., \& Szell, A.,
2006, ApJ, 648, 890.

\bibitem[MTJ]{MTJ} Merritt, D., Tremaine, S. \& Johnstone, D., 1989,
MNRAS,236,829.


\bibitem[Mutka 2009]{Mutka09} Mutka, P., 2009, Proceeding of \textit{Invisible
Universe}, Palais de l'UNESCO, Paris, ed. J-M Alimi.

\bibitem[Nakano \& Makino 1999]{NM99}Nakano, T., Makino, M., 1999,
ApJ, 525, L77.

\bibitem[NFW1996]{NFW}Navarro, J.F., Frenk, C.S., White, S.D.M.,
1996, ApJ, 462, 563.

\bibitem[Navarro et al., 2009]{diverse2009} Navarro, J.F., Ludlow,
A., Springel, V., Wang, J., Vogelsberger, M., White, S.D.M., Jenkins,
A., Frenk, C.S., \& Helmi, A., 2009, arxiv:0810.1522v2.

\bibitem[Peirani \& de Freitas Pacheo (2008)]{PFP2008}Peirani,S.
\& de Freitas Pacheo, J.A., 2008, Phys. Rev. D, 77 (6), 064023.

\bibitem[Peebles 1972]{P1972}Peebles, P.J.E., 1972, Gen.Rel.Grav.,
3, 63.

\bibitem[Quinlan et al., 1995]{Q1995}Quinlan, G.D., Hernquist, L.,
Sigurdsson, S., 1995, ApJ, 440, 554.

\bibitem[Stiavelli \& Bertin, 1985]{StiavelliBertin} Stiavelli, M.
\& Bertin, G., 1985, MNRAS,217,735.


%
{}

\bibitem[Young 1980]{Y1980}Young, P., 1980, ApJ, 242, 1232.%
{}

%
{}\end{thebibliography}

\end{document}